\documentclass[twocolumn,showpacs,preprintnumbers,amsmath,amssymb,prc]{revtex4}
\usepackage{graphicx}
\usepackage{dcolumn}% Align table columns on decimal point
\usepackage{bm}% bold math
\begin{document}
\title{Multiplicity derivative: A new signature of a first-order phase transition in intermediate energy heavy-ion collision}
\author{S. Mallik$^{1,2}$, G. Chaudhuri$^{1,2}$, P. Das$^{1,2}$ and S Das Gupta$^{3}$}
\affiliation{$^1$Physics Group, Variable Energy Cyclotron Centre, 1/AF Bidhan Nagar, Kolkata 700064, India}
\affiliation{$^2$Homi Bhabha National Institute, Training School Complex, Anushakti Nagar, Mumbai 400085, India}
\affiliation{$^3$Physics Department, McGill University, Montr{\'e}al, Canada H3A 2T8}
\begin{abstract}
Measurement of $M$, the total multiplicity, for central collision between comparable mass heavy ions can provide a signature for first-order phase transition.  The derivative of $M$ with respect to $E^*/A$ where $E^*$ is the excitation energy in the centre of mass and $A$ the total mass of the dissociating system is expected to go through maximum as a function of $E^*$.  Theoretical modeling shows that this is the energy where the specific heat $C_v$ maximizes which typically happens at the first-order phase transition.  The measurement of total $M$ is probably feasible in more than one laboratory.
\end{abstract}
\maketitle
{\bf {\it Introduction:-}}
In this article we suggest experiments which can provide evidence (or absence of evidence) for first order phase transition in intermediate energy heavy ion collisions.  Phase transitions occur in large systems and signatures of phase transition can be masked by finite sizes.  In nuclear physics the Coulomb interaction prevents formation of very large systems in the laboratory. In addition to limiting the size of nuclei, Coulomb effects further corrupt signatures of phase transition.  If finite size and Coulomb effects totally mask the signature of phase transition then no definite conclusions can be reached from the data. We suggest that the situation is not that ambiguous.\\
\indent
In a seminal paper Gulminelli and Chomaz pointed out that just the effect of finite size will cause bimodality
to appear in the mass distribution of composites for a first order transition \cite{Gulminelli}.  In heavy ion collisions (HIC) many
composites are produced.  Let us denote by $P_m(k)$ the probability that in the mass  distribution, the composite with
mass $k$ appears as the maximum mass.  One can plot $P_m(k)$ as a function of $k$.
In the case of first order phase transition, finite size produces two maxima at two different values of $k$.
The energy at which the two maxima achieve the same value defines the energy
of the bimodal point.  In thermodynamic models, instead of energy, the primary variable is the temperature.  We could talk
about the temperature of the bimodal point.  The range of energy or temperature where two maxima are seen can be called the
bimodal region.  It is a small region.  Bimodality has appeared in many calculations.  It was shown to appear in Boltzmann-Uehling-Uhlenbeck (BUU)
transport model of central collisions between two equal ions with Coulomb forces switched off \cite{Mallik14}.
It appeared in quantum molecular dynamics calculation \cite{Lefevre}.  It is seen in canonical thermodynamic model \cite{Chaudhuri1,Chaudhuri2,Mallik13}.\\
\indent
The question we ask is as follows: if the corruptive effects of Coulomb interaction is so strong that bimodality is destroyed is there any other observable that points to vestiges of a first order phase transition?  Our answer is yes.  We use the canonical thermodynamic model (CTM) \cite{Das} to establish our claim.  But we need first to turn to bimodality in CTM without and with Coulomb interaction.\\
\indent
Two different microcanonical versions employing similar physics as CTM are Statistical Multifragmentation Model (SMM) by the Copenhagen group \cite{Bondorf1} and Microcanonical Metropolis Monte Carlo (MMMC) by the Berlin group \cite{Gross}. All these models \cite{Das,Bondorf1,Gross} were very successfully used to fit many data in HIC. Results from CTM and SMM have been found to be very close \cite{Chaudhuri_plb}. Here we use CTM. We will skip all calculational details of CTM as they can be found in many places. Composites carry charge and the long range coulomb interactions between composites are included in Wigner-Seitz approximation \cite{Bondorf1} in SMM and also adopted in CTM. We will use results from a previous calculation and in particular, Fig. 1 of Ref. \cite{Chaudhuri2}. The example studied dissociation of a system with $N$ and $Z$ equal to 75.  The coulomb effects were studied by varying the strength of coulomb interaction using a
multiplicative factor $x_c$.  $x_c$=0 means no coulomb interaction, $x_c$=1 means the actual strength of coulomb force. An intermediate value of $x_c$ means a reduced value of coulomb interaction. The lesson that we learn from that work is this. For $x_c$=0 bimodality appears.  In addition the specific heat $c_v$ hits a maximum value at the bimodal point.  For small values of  $x_c$ bimodality region is shrinking and the maximum value of $c_v$ is close to the bimodal temperature but not identical.  Bimodality disappears before reaching $x_c$=1 but the usual behaviour of $c_v$ reaching a maximum at phase transition temperature continues until $x_c$=1.  So if we could measure the $c_v$ we would see vestiges of first order phase transition even with the usual coulomb force. But since measuring $c_v$ is not a practical suggestion, is there some other measurable quantity that also maximises when $c_v$ does? Theoretical modelling predicts that the derivative of total multiplicity with respect to temperature displays a maximum which coincides with the maximum of $c_v$.  This is shown in the next section.  Since temperature increases with increasing beam energy the maximum can be located in experiments.\\
\begin{figure}
\includegraphics[width=6.5cm,keepaspectratio=true]{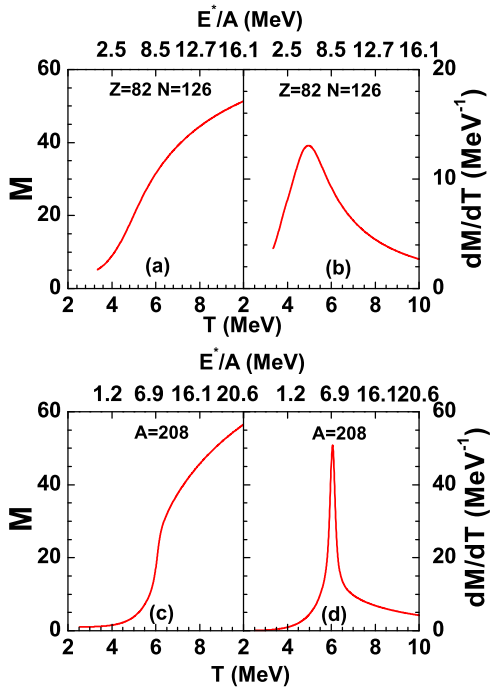}
\caption{(Color online) Variation of multiplicity $M$ (left panels) and d$M$/d$T$ (right panels) with temperature (bottom $x$-axes) and excitation per nucleon (top $x$-axes)from CTM calculation  for fragmenting systems having $Z$=82 and $N$=126 (top panels). Bottom panels represent the same but for hypothetical system of one kind of particle with no coulomb interaction but the same mass number ($A$=208). $E^*$ is $E-E_0$ where $E_0$ is the ground state energy of the dissociating system in the liquid drop model whose parameters are given in Ref. \cite{Das}}
\end{figure}
\begin{figure}
\includegraphics[width=6.5cm,keepaspectratio=true]{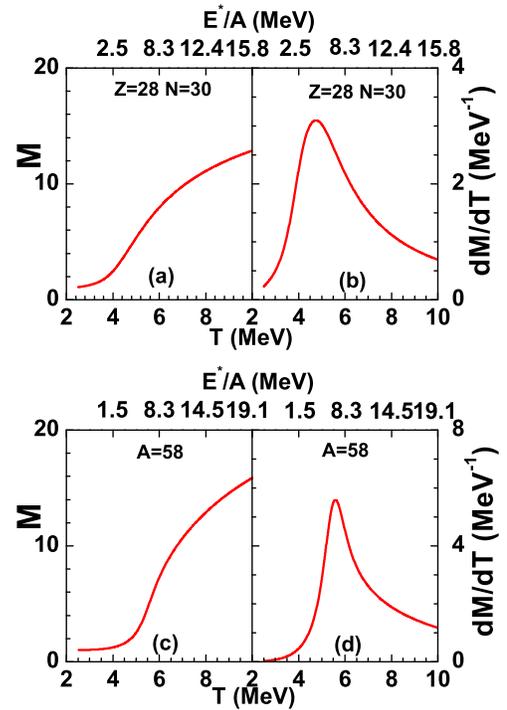}
\caption{(Color online) Same as Fig. 1 but the fragmenting systems are $Z$=28 and $N$=30 (top panels) and $A$=58 (bottom panels)}
\end{figure}
{\bf {\it Results:-}}
In central collisions of nearly equal mass ions one can measure with 4$\pi$ detectors the total multiplicity $M=\sum M_a$.
Here $a$ denotes the mass numbers of composites.  In
CTM the derivative of $M$ with $T$ as a function of $T$ is seen to have a maximum.  Fig.1 (left panel) shows the total multiplicity for fragmenting system having proton number ($Z$)=82 and neutron number ($N$)=126 and and its derivative $dM/dT$ (the right panel). Results for both real nuclei and the one for one kind of particles have been displayed in order to emphasize the effects of Coulomb interaction. The rise and the peak are much sharper in absence of Coulomb interaction clearly indicating the role of the long range interaction. As the system size decreases (Fig.2), the features become less sharp as in $Z$=28 and $N$=30. The peak in $dM/dT$ coincides with the maximum of  specific heat at constant volume $C_v$  as a function of temperature and this is seen in Fig 3 and 4 for $Z$=82, $N$=126 and $Z$=28, $N$=30 respectively.
Of course experiments do not give
$T$ directly but a plot against $E^*/A$ will also show a nearly coincident maximum (see Fig 1 and 2).
 The peak in $C_v$
 is a signature of first order phase transition. In $dM/dT$, we have the peak coinciding with that of $C_v$ and hence  we are proposing it as
a new method for testing the occurrence of first order phase transition in HIC.  Even where bimodality develops, it may be easier to
locate the position of the maximum in the derivative of $M$ since the bimodal region is very narrow.\\
\begin{figure}
\includegraphics[width=8cm,keepaspectratio=true]{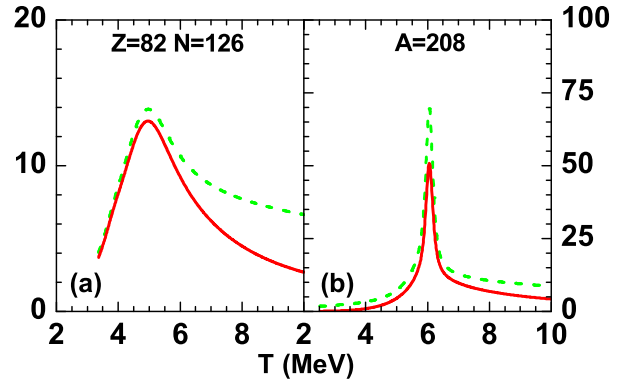}
\caption{(Color online) Variation of d$M$/d$T$ (red solid lines) and $C_v$ (green dashed lines) with temperature from CTM for fragmenting systems having $Z$=82 and $N$=126 (left panel) and for hypothetical system of one kind of particle with no coulomb interaction of mass number $A$=208. To draw d$M$/d$T$ and $C_v$ in the same scale, $C_v$ is normalised by a factor of 1$/$50.}
\end{figure}
\begin{figure}
\includegraphics[width=8cm,keepaspectratio=true]{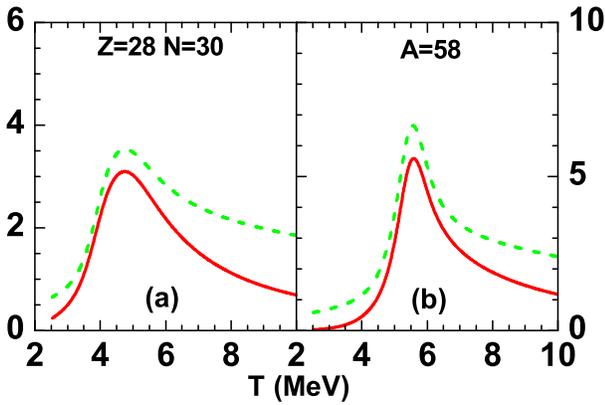}
\caption{(Color online) Same as Fig. 3 but the fragmenting systems are $Z$=28 and $N$=30 (left panel) and $A$=58 (right panel)}
\end{figure}
\begin{figure}
\includegraphics[width=5.5cm,keepaspectratio=true]{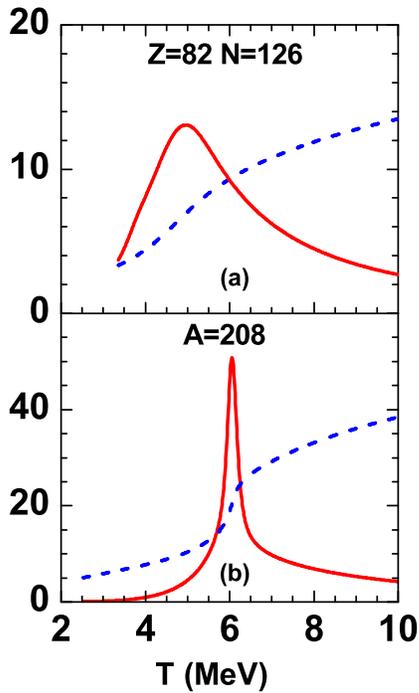}
\caption{(Color online) Variation of entropy (blue dashed lines) and d$M$/d$T$ (red solid lines) with temperature from CTM for fragmenting systems having $Z$=82 and $N$=126 (top panel) and for hypothetical system of one kind of particle with no coulomb interaction of mass number $A$=208 (bottom panel). To draw $S$ and d$M$/d$T$ in the same scale, $S$ is normalised by a factor of 1$/$20 for $Z$=82 and $N$=126 system and 1$/$50 for hypothetical system of one kind of particle.}
\end{figure}

\begin{figure}
\includegraphics[width=9cm,keepaspectratio=true]{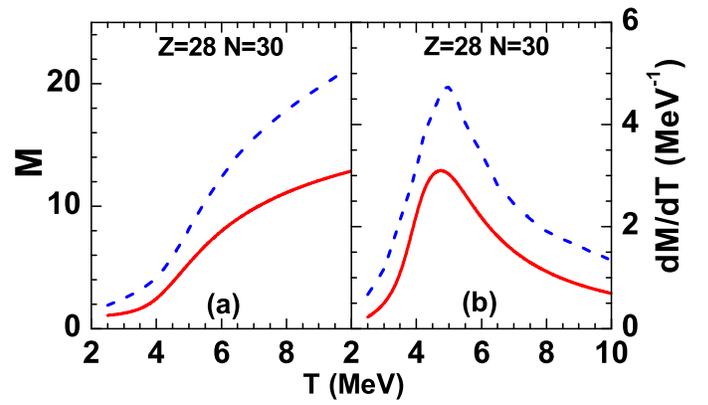}
\caption{(Color online) Effect of secondary decay on on $M$ (left panel) and d$M$/d$T$ (right panel) for fragmenting systems having $Z$=28 and $N$=30. Red solid lines show the results after the multifragmentation stage (calculated from CTM) where as blue dashed lines represent the results after secondary decay of the excited fragments.}
\end{figure}
\begin{figure}[b]
\includegraphics[width=8cm,keepaspectratio=true]{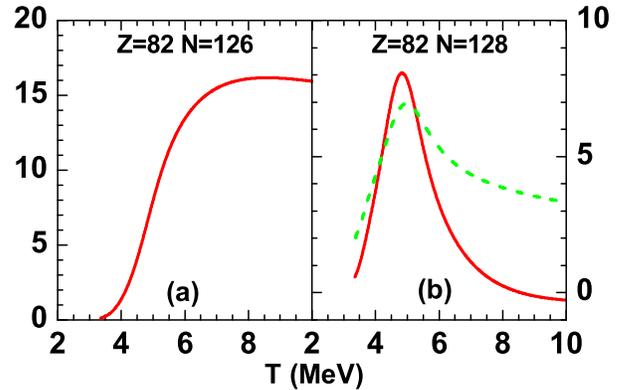}
\caption{(Color online) Variation of intermediate mass fragment (IMF) multiplicity $M_{IMF}$ (left panels) and first order derivative of IMF multiplicity  d$M_{IMF}$/d$T$ (right panels) with temperature from CTM calculation  for fragmenting systems having $Z$=82 and $N$=126. Variation of $C_v$ with temperature ($T$) is shown by green dashed line in right panel. To draw d$M_{IMF}$/d$T$ and $C_v$ in the same scale, $C_v$ is normalised by a factor of 1$/$100. }
\end{figure}
\indent
It is  also worth mentioning that near the maximum of $dM/dT$ the entropy of the dissociating system makes a higher jump than is seen
far from it.  This is also shown in Fig. 5. For the hypothetical (one particle) system, the increase in entropy near the maximum of $dM/dT$ is much more pronounced (lower panel), while the Coulomb interaction effect smears the rise in the real system (upper panel). \\
\indent
It is well known that composites from CTM are excited and hence will undergo sequential two body decay \cite{Mallik1} which will change the total multiplicity. We have examined this and found that this will not alter our conclusions. In fact, sequential decay makes the peak in dM/dT sharper. This is shown in Fig. 6.\\
\indent
Lastly we have examined the features of intermediate mass fragments (composites with charge $3{\le}z{\le}20$) and it is observed that
similar behaviour is also displayed by $M_{IMF}$ and its derivative as shown in Fig. 7. It was shown  earlier for an
 idealised system with one kind of particles that
there is a dramatic increase in $M_{IMF}$ in a short temperature interval \cite{Dasgupta_Phase_transition}. For  the sake of completeness, we have
also shown here how $M_{IMF}$ behaves with temperature with Coulomb interaction included.
The peak in the  derivative does not exactly coincide with that of $C_v$ and this is expected since in $M_{IMF}$  all the composites and nucleons are not included which are used in the calculation of $C_v$. $M_{IMF}$ is also an important experimental observable which is measured in many situations \cite{Peaslee,Ogilvie,Tsang} instead of the total multiplicity $M$.  However prescription of considering full $M$ and its derivative is more precise in locating the position
of the maximum of $c_v$ which signifies that we are at first order phase transition.\\
{\bf {\it Discussions:-}}
Establishing evidence for phase transition in nuclear matter from data obtained from intermediate
energy heavy ion collision has attracted much attention in the last twenty years.  Here we have used
measurable $dM/dE$ and $dM_{IMF}/dE$ as evidence for first order phase transition
should a maximum be seen.  The answer is unambiguous: it is either yes or no.  Most past 
investigations have suffered from ambiguity.  An example was trying to fit an individual $M_a$
to $a^{-\tau}f(a^{\sigma}(T-T_c))$ \cite{Elliot1,Elliot2,Scharenberg,Fisher,Stauffer}.
Equally acceptable but quite approximate fits were found with very different models so no
conclusions could be made.  One model that predicted first order phase transition was the lattice gas model \cite{Pan1} but the property of $M$ was not investigated.  It will be interesting to pursue
that.
\indent


\begin{thebibliography}{99}
\bibitem{Gulminelli} F. Gulminelli and Ph. Chomaz, Phys.Rev. C {\bf 71}, 054607 (2005).
\bibitem{Mallik14} S. Mallik, S. Das Gupta and G. Chaudhuri, Phys. Rev. C {\bf 93}, 041603 (2016)(R).
\bibitem{Lefevre} A. Lefevre and J. Aichelin, Phys. Rev. Lett. {\bf 100}, 042701 (2008).
\bibitem{Chaudhuri1} G. Chaudhuri and S. Das Gupta, Phys. Rev. C {\bf 76}, 014619 (2007).
\bibitem{Chaudhuri2} G. Chaudhuri, S. Das Gupta and F. Gulminelli, Nucl. Phys A {\bf 815}, 89 (2009).
\bibitem{Mallik13} S. Mallik, F. Gulminelli and G. Chaudhuri, Phys. Rev. C {\bf 92}, 064605 (2015).
\bibitem{Das} C. B. Das, S. Das Gupta, W. G. Lynch, A.Z. Mekjian and M. B. Tsang, Phys. Rep. {\bf 406}, 1 (2005).
\bibitem{Bondorf1} J. P. Bondorf, A. S. Botvina, A. S. Iljinov, I. N. Mishustin and K. Sneppen, Phys. Rep. {\bf 257}, 133 (1995).
\bibitem{Gross} D. H. E. Gross, Phys. Rep. {\bf 279}, 119 (1997).
\bibitem{Chaudhuri_plb} A. S. Botvina, G. Chaudhuri, S Das Gupta and I. N. Mishustin, Phys. Lett B {\bf 668}, 414 (2008).
\bibitem{Mallik1} G. Chaudhuri and S. Mallik, Nucl. Phys. {\bf A 849}, 190 (2011).
\bibitem{Dasgupta_Phase_transition} S. Das Gupta, A. Z. Mekjian and M. B. Tsang, {\it Advances in Nuclear Physics}, Vol. 26, 89 (2001) edited by J. W. Negele and E. Vogt, Plenum Publishers, New York.
\bibitem{Elliot1} J. B. Elliott et al., Phys. Lett. B {\bf 381}, 35 (1996).
\bibitem{Elliot2} J. B. Elliott et al., Phys. Lett B {\bf 418}, 34 (1998).
\bibitem{Scharenberg} R. P. Scharenberg et al.,phys. Rev. C {\bf 64}, 054602 (2001).
\bibitem{Fisher} M. E. Fisher Physics {\bf 3},255 (1965).
\bibitem{Stauffer} D. Stauffer, A. Aharoni, {\it Introduction to percolation theory}, Taylor and Francis, Washington D. C .1992 (chapter 2).
\bibitem{Peaslee} G. F. Peaslee et al., Phys. Rev. {\bf C 49}, 2271 (1994)(R).
\bibitem{Ogilvie} C. A. Ogilvie et al., Phys. Rev. Lett. {\bf  67}, 1214 (1991).
\bibitem{Tsang} M. B. Tsang et al., Phys. Rev. Lett. {\bf  71}, 1502 (1993).
\bibitem{Pan1} J. Pan, S. Das Gupta and M. Grant, Phys. Rev. Lett {\bf 80}, 1182 (1998).

%\bibitem{Borderie2} B. Borderie and M. F. Rivet, Prog. Part. Nucl. Phys. {\bf 61}, 551 (2008).
%\bibitem{Gross_phase_transition} D. H. E. Gross, Prog. Part. Nucl. Phys. {\bf 30}, 155 (1993).
%\bibitem{Bauer} W.Bauer,Phys. Rev. C {\bf 38},1297 (1988).
%\bibitem{Campi} X. Campi,Phys. Lett. B {\bf 208},351 (1988).
%\bibitem{Pan1} J. Pan, S. Das Gupta and M. Grant, Phys. Rev. Lett {\bf 80}, 1182 (1998).
%\bibitem{Hufner} J.Hufner and D. Mukhopadhyay, Phys. Lett. B {\bf 173}, 373 ((1986)
%\bibitem{Oddershede} L. Oddefshede, P. Dimon, J. Bohr, Phys. Rev. Lett. {\bf 71}, 3107 (1993)
%\bibitem{Das1} C. B. Das, S. Das Gupta, W. G. Lynch, A. Z. Mekjian and M. B. Tsang, Phys. Rep. {\bf 406}, 1 (2005).
%\bibitem{Bondorf} J. P. Bondorf et. al., Phys. Rep. {\bf 257}, 133 (1995).
%\bibitem{Dasgupta1} S. Das Gupta and A. Z. Mekjian, Phys. Rev. C {\bf 57}, 1361 (1998).


\end{thebibliography}
\end{document}